# Data Fusion and Machine Learning Integration for Transformer Loss of Life Estimation

Mohsen Mahoor, Amin Khodaei
Dept. of Electrical and Computer Engineering
University of Denver
Denver, CO, USA
Mohsen.Mahoor@du.edu, Amin.Khodaei@du.edu

*Abstract*— **Rapid growth of machine learning methodologies and their applications offer new opportunity for improved transformer asset management. Accordingly, power system operators are currently looking for data-driven methods to make better-informed decisions in terms of network management**. **In this paper, machine learning and data fusion techniques are integrated to estimate transformer loss of life. Using IEEE Std. C57.91-2011, a data synthesis process is proposed based on hourly transformer loading and ambient temperature values. This synthesized data is employed to estimate transformer loss of life by using Adaptive Network-Based Fuzzy Inference System (ANFIS) and Radial Basis Function (RBF) network, which are further fused together with the objective of improving the estimation accuracy. Among various data fusion techniques, Ordered Weighted Averaging (OWA) and sequential Kalman filter are selected to fuse the output results of the estimated ANFIS and RBF. Simulation results demonstrate the merit and the effectiveness of the proposed method.**

NOMENCLATURE

*Parameters:*

| | |
|---|---|
| $A$ | Transition matrix. |
| $B$ | Control matrix. |
| $C_1 / C_2$ | Weight factors. |
| $H$ | Observation matrix. |
| $I$ | Identity matrix. |
| $K$ | Kalman gain. |
| $P$ | Error covariance |
| $Q$ | Process noise covariance. |
| $E$ | Measurement noise covariance. |
| $S$ | Number of train datasets. |
| $u$ | Exogenous control input. |
| $Z$ | A new value estimated either from the ANFIS or the RBF method. |
| $Y$ | Target values for transformer loss of life. |
| $\hat{Y}$ | ANFIS estimated values for transformer loss of life. |
| $\hat{O}$ | RBF estimated values for transformer loss of life. |
| $\hat{x}$ | State estimate. |
| $F_{AA}$ | Aging acceleration factor of insulation. |
| $F_{AA,n}$ | Aging acceleration factor during $\Delta t_n$. |
| $F_{EQA}$ | Equivalent aging factor. |
| $m/n$ | An empirically derived exponent used to calculate the variation of $\Delta\theta_H / \Delta\theta_{TO}$ with changes in load. |
| $R$ | The ratio of load loss. |
| $\Delta t_n$ | Time interval. |
| $\theta_H$ | Winding hottest-spot temperature (°C). |
| $\theta_A$ | Ambient temperature (°C). |
| $\Delta\theta_H$ | Winding hottest-spot rise over top-oil temperature (°C). |
| $\Delta\theta_{TO}$ | Top-oil rise over ambient temperature (°C). |

*Subscripts:*

| | |
|---|---|
| $H$ | Winding hottest-spot |
| $i/U$ | Initial/Ultimate |
| $R$ | Rated |
| $TO$ | Top oil |
| $w$ | Winding |
| $s$ | Index for training datasets. |

## I. INTRODUCTION

ASSET MANAGEMENT, as one of the key duties of electric utility companies, is frequently performed to ensure power system reliability and security. Component maintenance and repair services are considered as the significant elements of the asset management, which aim at minimizing component failures and power outages. Given the fact that the electricity infrastructures in the US is aging (mainly constructed in 1950s and 60s), and the consumers' demand for a reliable and secure service is growing, this problem has attracted more attention in recent years [1]-[5].

Among power system components, the distribution transformer is one of the fundamental and pivotal elements that its maintenance and management need to be continuously investigated by electric utility companies. Distribution transformers play a vital role in ensuring a reliable power supply as their failure will commonly result in unplanned power outages. Moreover, transformers not only are considered as a cost-intensive component in power systems, but also their maintenance and repair services are labor-intensive and time-consuming [1], [2]. Various approaches are carried out for transformer asset management, including condition monitoring, online monitoring, routine diagnostic, scheduled maintenance, and condition based maintenance, to name a few [6]. Considering that the transformer insulation is more likely subject to failure than other parts, the transformer lifetime is commonly investigated based on the condition of the transformer insulation. In this case, the transformer's internal temperature plays a decisive role in effecting the transformer aging, specifically, the internal temperature of the hottest-spot. The hottest-spot temperature is a function of transformer loading and ambient temperature [7].

The IEEE Std. C57.91-2011 Guide for Loading Mineral-Oil-Immersed Transformers comprehensively presents a series of functions to calculate the transformer loss of life [8]. The study in [9] provides a framework in which on the basis of the IEEE standard and by using sensory data, transformer lifetime is estimated in a dynamic manner. In [10], estimated hourly load and ambient temperature are fed to the IEEE standard in order to evaluate the failure time of the transformer insulation as well as estimating its remaining life. By utilizing the available data for transformer loss of life and based on the discussed IEEE standard, a risk-based probabilistic approach is proposed to assess transformer health in [11]. A joint midterm and short-term maintenance scheduler for power transformer asset management is presented in [5] based on failure rate modeling and N-1 reliability criterion. The study in [12] uses both data quality control and data-set screening approaches to demonstrate that the reliability of the IEEE standard is increased. In [13], first, by using MATLAB, transformer loss of life is estimated, and then the estimated values are validated through comparing with experimental data. The strength of survival data on the accuracy of transformer statistical lifetime models is examined in [14] via Monte Carlo simulations, where the results demonstrate that the accuracy of models can be enhanced by taking the survival data into consideration. Leveraging historic data and employing captured data from online sensor measurements, an intelligent scheme for condition monitoring and assessment of power transformers is proposed in [2]. Within the proposed intelligent scheme, several types of signal processing and pattern recognition approaches are applied for identifying transformer faults, and also processing data and information.

Given that a significant amount of data can be collected from sensors installed in transformers, machine learning methods can be of value in estimating transformer lifetime. A machine learning-based study with the goal of estimating transformer loss of life is proposed in [4]. By leveraging the historical data of transformer loading and ambient temperature, various machine learning methods, including Adaptive Network-Based Fuzzy Inference System (ANFIS), Multi-Layer Perceptron (MLP) network and Radial Basis Function (RBF) network are employed to accurately estimate the transformer loss of life. Authors in [15] utilize a fuzzy modeling system for transformer asset management. An artificial neural network model for predicting top oil temperature in transformer is used in [16]. A naïve thermal model to estimate transformer lifetime and transformer replacement time on the basis of an evolutionary algorithm, here genetic program and by using experimental data, is presented in [17].

To the best of the authors' knowledge, the existing literature in this research area lacks studies on data-driven methodologies, such as machine learning and data fusion, for transformer lifetime assessment. The primary objective in this paper is to integrate data fusion and machine learning techniques for providing a more accurate and reliable estimation of transformer loss of life. Various types of machine learning methods to estimate the transformer loss of life, as proposed in the companion paper [4], set the stage for using data fusion techniques, and thus call for additional studies. In general, all tasks that demand any type of estimation from multiple sources can reap the benefit of using data fusion techniques. The following well-known definition of data fusion is provided in [18]: "data fusion techniques combine data from multiple sensors and related information from associated databases to achieve improved accuracy and more specific inferences than could be achieved by the use of a single sensor alone." Two types of data fusion techniques, including Ordered Weighted Averaging (OWA) and Kalman filter are presented and studied in this paper. By leveraging these data fusion techniques, two premier outputs estimated by machine learning methods [4], are fused together to improve the transformer loss of life estimation. Comparison between the proposed data fusion techniques, i.e., OWA and Kalman filer, is further provided in this work.

The remainder of this paper is organized as follows. Section II describes the data synthesis process based on the IEEE standard. Machine learning and data fusion techniques are explained in Section III. Section IV presents numerical simulations and analyses to show the effectiveness of the proposed data fusion techniques. Conclusions drawn from this paper are presented in Section V.

II. DATA SYNTHESIS BASED ON THE IEEE STD. C57.91-2011

The IEEE Std. C57.91-2011 provides a series of nonlinear functions, mainly based on the winding hottest-spot temperature, to calculate the transformer loss of life. Given the fact that the temperature does not have a uniform distribution in transformer, the hottest-spot is taken into account for calculations. It is interesting to note that the Arrhenius' chemical reaction rate theory is the origin of the IEEE standard experimental equations for quantifying the transformer loss of life. The per-unit life of transformer is defined in (1):

$$\text{Per unit life} = A \exp(\frac{B}{\theta_H + 273}), \quad (1)$$

where $A$ and $B$ are per unit constant and the aging rate, respectively. Equation (1) is the starting point to find Aging Acceleration Factor (AAF) for a given winding hottest-spot temperature, as defined in (2):

$$F_{AA} = \exp(\frac{15000}{383} - \frac{15000}{\theta_H + 273}). \quad (2)$$

Accordingly, (2) is used to calculate the equivalent aging of transformer (3):

$$F_{EQA} = \sum_{n=1}^{N} F_{AA_n} \Delta t_n \bigg/ \sum_{n=1}^{N} \Delta t_n, \quad (3)$$

where $\Delta t_n$ is the time interval, $n$ is the time interval index and $N$ is the total number of time intervals. The percentage of insulation loss of life can therefore be defined as follows:

$$LOL(\%) = \frac{F_{EQA} \times t \times 100}{\text{Normal insulation life}}. \quad (4)$$

The hottest-spot temperature is calculated as:

$$\theta_H = \theta_A + \Delta\theta_{TO} + \Delta\theta_H, \quad (5)$$

where $\theta_A$ represents ambient temperature, $\Delta\theta_{TO}$ is top-oil rise over ambient temperature, and $\Delta\theta_H$ is the winding hottest-spot rise over top-oil temperature. $\Delta\theta_{TO}$ and $\Delta\theta_H$ are respectively characterized by (6) and (7):

$$\Delta\theta_{TO} = (\Delta\theta_{TO,U} - \Delta\theta_{TO,i})(1 - \exp(-\frac{1}{\tau_{TO}})) + \Delta\theta_{TO,i}, \quad (6)$$

$$\Delta\theta_H = (\Delta\theta_{H,U} - \Delta\theta_{H,i})(1 - \exp(-\frac{t}{\tau_w})) + \Delta\theta_{H,i}. \quad (7)$$

Finally, the initial and ultimate values for $\Delta\theta_{TO}$ and $\Delta\theta_H$ are calculated via (8)-(11) as follows:

$$\Delta\theta_{TO,i} = \Delta\theta_{TO,R}(\frac{K_i^2 R + 1}{R + 1})^n, \quad (8)$$

$$\Delta\theta_{TO,U} = \Delta\theta_{TO,R}(\frac{K_U^2 R + 1}{R + 1})^n, \quad (9)$$

$$\Delta\theta_{H,i} = \Delta\theta_{H,R} K_i^{2m}, \quad (10)$$

$$\Delta\theta_{H,U} = \Delta\theta_{H,R} K_U^{2m}. \quad (11)$$

Considering (1)-(11), obtained from the IEEE standard, it can be seen that the transformer loss of life is a function of both transformer loading and ambient temperature. In other words, as shown in Fig. 1, by plugging the hourly values of transformer loading and ambient temperature into the above-mentioned equations, the hourly transformer loss of life could be calculated. This process is called data synthesis in which the hourly value of the transformer loss of life is synthesized on the basis of this IEEE standard. The synthesized data is utilized to be employed in machine learning methods and data fusion techniques for estimating the transformer loss of life.

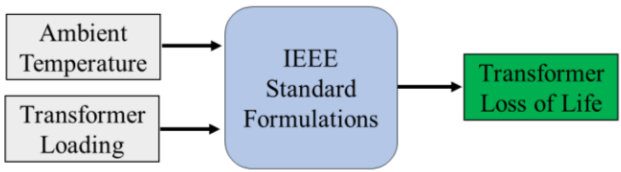

Fig. 1. Data synthesis process based on the IEEE standard.

III. MACHINE LEARNING AND DATA FUSION

An overview of machine learning methods and data fusion techniques for estimating transformer loss of life is presented in this section. Data fusion techniques are utilized to improve the machine learning estimated values of the transformer loss of life. In fact, data fusion is used to fuse the outputs of the ANFIS and the RBF methods in such a way that the estimated transformer loss of life becomes more accurate. In what follows, machine learning methods to estimate the transformer loss of life are provided, then two various kinds of data fusion techniques, including Ordered Weighted Averaging (OWA) and sequential Kalman filter, are introduced with the goal of integrating machine learning and data fusion.

*A. Machine Learning*

Machine learning is an intelligent method to solve nonlinear estimation and classification problems [19]. Various data-driven machine learning methods, including but not limited to ANFIS, RBF and MLP, can be considered as suitable candidates for solving the estimation problems. Without addressing the details of these machine learning methods, and by referring to the companion paper [4], the transformer loss of life is estimated using these three methods, as shown in Fig. 2(a). Each of these machine learning methods have different performances, which are quantified by two measures: Mean Square Error (MSE) and coefficient of determination ($R^2$). It should be noted that these performance measures, i.e., MSE and $R^2$, are applicable to data fusion techniques as well. Two data fusion techniques are presented here to combine the two aforementioned machine learning methods, i.e., the ANFIS and the RBF, with the objective of improving the accuracy of the transformer loss of life estimation.

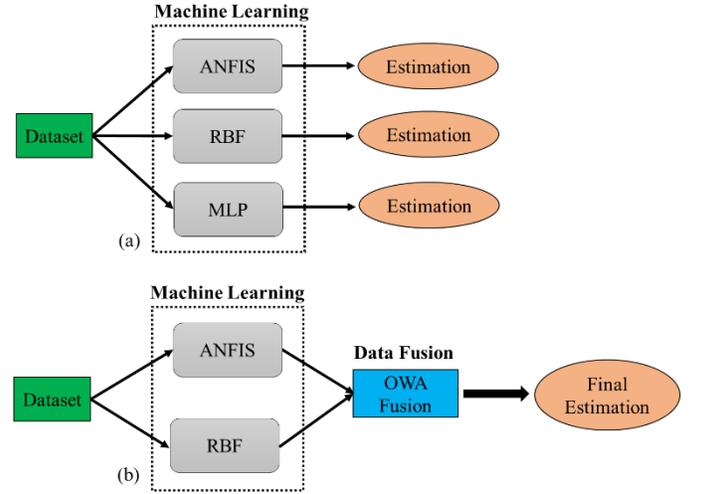

Fig. 2. Transformer loss of life estimation by using (a) machine learning, (b) machine learning and OWA fusion.

*B. Ordered Weighted Averaging-Based Data Fusion*

OWA operator, as one of the most popular data fusion techniques, has been introduced in [20]. OWA is utilized to incorporate the output results of the estimated ANFIS and RBF methods, as shown in Fig. 2(b). To this end, the objective function that should be minimized is as follows:

$$\min MSE = \frac{1}{S}\sum_{s=1}^{S}\left(C_1 \times \hat{Y}_S + C_2 \times \hat{O}_S - Y_S\right)^2, \quad (12)$$

$$C_1, C_2 \in [0,1] \quad and \quad C_1 + C_2 = 1, \quad (13)$$

where $C_1$ and $C_2$ are weight factors corresponding to the ANFIS and RBF, respectively. $Y_s$ is the target value for the transformer loss of life, and $S$ is the number of training dataset. Moreover, $\hat{Y}_s$ and $\hat{O}_s$ are respectively the ANFIS and the RBF estimated values of the transformer loss of life.

Genetic Algorithm (GA) is employed in order to obtain the optimal contribution of each machine learning method to build the OWA-based data fusion [21]. Accordingly, GA determines the optimal weight factors, i.e., $C_1$ and $C_2$, which aims at minimizing the objective function. After running GA, the optimized weight factors are acquired to be employed in the test dataset to yield the final estimation.

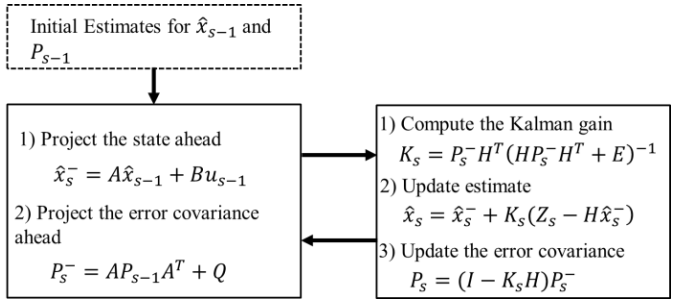

Fig. 4. Kalman filter algorithm.

## C. Kalman Filter-Based Data Fusion

The Kalman filter was developed by R. Kalman. In 1960 his well-known paper [22] was published with the goal of unknown system state estimation via filtering behavior. Generally speaking, Kalman filters encompass a number of types and topologies depending on use and application. In this section, on the basis of the Kalman filter, a sequential processing technique is developed for the purpose of data fusion. Fig. 3 demonstrates an overview of the sequential update architecture for data fusion using the Kalman filter.

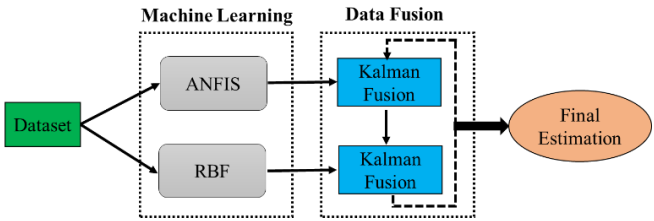

Fig. 3. Architecture of the sequential Kalman filter fusion.

The recursive equations of the Kalman filter are shown in Fig. 4. At each sample point, the algorithm projects both the state estimate, i.e., $x_s$, and the error covariance, i.e., $P_s$. In the second stage, the Kalman gain, i.e., $k_s$, is computed. Then, by incorporating a new value, i.e., $z_s$, the improved estimate is updated. Finally, the error covariance is updated. It is assumed that the process noise covariance, i.e., $Q$, and the measurement noise covariance, i.e., $E$, are not changing with each sample point, so that they both are considered as constant matrices. Noted that $u_s$ and $H$ are exogenous control input and observation matrix, respectively. In addition, $A$ and $B$ are respectively transition and control matrices. More mathematical details and explanations can be found in [23].

## IV. NUMERICAL SIMULATIONS

In this section, the performance of the machine learning methods and the data fusion techniques for estimating the transformer loss of life is evaluated. In this regard, the required data is synthesized on the basis of the mentioned IEEE standard. More details of the data synthesis process are not reported in this paper, but available in the companion paper [4]. The following cases are studied to investigate the performance of integration of the machine learning and data fusion techniques for estimating the transformer loss of life.

Case 0:  Transformer loss of life estimation using machine learning methods.
Case 1:  Transformer loss of life estimation using OWA-based data fusion.
Case 2:  Transformer loss of life estimation using Kalman filter-based data fusion.

**Case 0**: Three machine learning methods, including ANFIS, RBF and MLP, are applied to the synthesized data to estimate the transformer loss of life. As reported with details in [4], among these three methods, two of them (ANFIS and RBF) outperform the other one (MLP) in terms of having lower MSE and higher $R^2$, so that these two superior methods are selected to be fused together, as will be carried out in Cases 1 and 2. The MSE and $R^2$ in the ANFIS method, applied in the test datasets, are respectively calculated as $2.946 \times 10^{-10}$ and 0.96. For the RBF method, $4.124 \times 10^{-10}$ and 0.89 are the best obtained values for the MSE and $R^2$, respectively.

**Case 1**: The OWA-based data fusion is employed in this case to combine the two selected machine learning methods of Case 0. The proposed OWA operator is modeled in MATLAB for fusing the hourly estimated values of the transformer loss of life. After running the GA, the optimized weight factors, i.e., $C_1$ and $C_2$, for fusing the output of the ANFIS and RBF are obtained as 0.9 and 0.1, respectively. The MSE and $R^2$ using the OWA-based data fusion are $2.832 \times 10^{-10}$ and 0.97, respectively. This case advocates the fact that by leveraging the OWA-based data fusion technique, the accuracy of the results is improved. In fact, compared to each of the machine learning methods in Case 0, this data fusing technique leads to lower MSE and higher $R^2$ for estimating the transformer loss of life.

**Case 2:** The Kalman filter-based data fusion is used in this case. The estimated output results of the ANFIS and RBF are fused in a sequential manner using the Kalman filter equations to achieve better performance measures. It is worth to mention that in the proposed Kalman filter algorithm, both $A$ and $H$ are equal to 1, and $B$ is 0. Moreover, $z_z$ is an estimated value achieved either from the ANFIS or the RBF method. After running the simulation, the values of MSE and $R^2$ are calculated as $2.389 \times 10^{-10}$ and 0.99, respectively, which outperforms the corresponding values in Cases 0 and 1. Fig. 5 compares the Kalman filter-fused values of the transformer loss of life with the actual ones, obtained from the data synthesis process, as well as the error (the difference between these two values). It should be noted that Fig. 5 is depicted only for 50 samples of the test datasets to provide a better visual comparison.

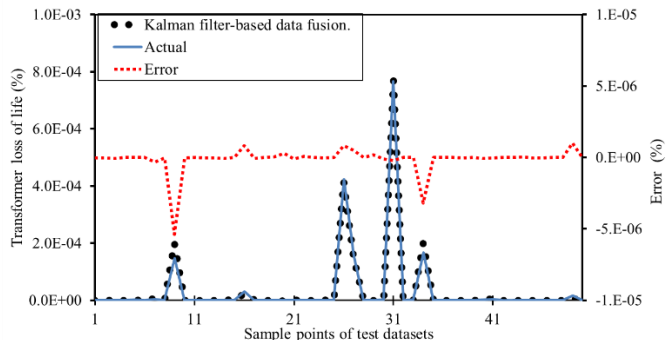

Fig. 5. Comparison between Kalman filter-fused values of the transformer loss of life with the actual ones.

The obtained results from these three case studies are ranked based on the two performance measures (MSE and $R^2$), and tabulated in Table I. As the table demonstrates, integrating machine learning methods and data fusion techniques enhance the accuracy of the transformer loss of life estimation. A comparison between Cases 1 and 2 advocates that the Kalman-filter-based data fusion technique surpasses the OWA-based one in terms of performance measures. It should be noted that as the simulations are carried out offline, the computation times are of no importance, thus not listed here. Taking all the results into consideration, it is admitted that incorporating the machine learning methods and the data fusion techniques boosts the accuracy of the transformer loss of life estimation.

TABLE I
COMPARISON OF THE MACHINE LEARNING METHODS AND DATA FUSION TECHNIQUES FOR ESTIMATING THE TRANSFORMER LOSS OF LIFE

|  |  | MSE | $R^2$ | Rank |
|---|---|---|---|---|
| **Machine Learning** | ANFIS | $2.946 \times 10^{-10}$ | 0.96 | 3 |
|  | RBF | $4.124 \times 10^{-10}$ | 0.89 | 4 |
| **Data Fusion** | OWA | $2.832 \times 10^{-10}$ | 0.97 | 2 |
|  | Kalman Filter | $2.389 \times 10^{-10}$ | 0.99 | 1 |

## V. CONCLUSIONS

Transformer maintenance and repair service has always been one of the priorities of power system operators, as transformer failure causes unplanned outages and can negatively impact power system reliability. A methodology to obtain a low-error estimate of transformer loss of life was proposed in this paper, leveraging an integrated machine learning and data fusion technique. The IEEE Std. C57.91-2011 was used to synthesize data, followed by two machine learning methods, including the ANFIS and RBF, to estimate the transformer loss of life. Then, by leveraging the OWA operator and the Kalman filter, the estimated results of these two machine learning methods were fused together to obtain a more accurate estimate. The proposed Kalman filter-based data fusion technique outperforms OWA as well as individual machine learning methods in terms of the MSE and $R^2$.